\documentclass[useAMS,usenatbib]{mn2e}

\usepackage{times}
\usepackage{graphicx}
\newcommand{\asymerr}[2]{\tiny\makebox[0pt][l]{\raisebox{-0.75ex}{$-#1$}}%
   \raisebox{+1.25ex}{\makebox[0pt][l]{$+$}\phantom{$-$}$#2$}}

\title[The evolution of the brightest cluster galaxies since $z\sim1$]{
  The
  evolution of the brightest cluster galaxies since $z\sim1$
  from the ESO Distant Cluster Survey (EDisCS)}
  \author[I.~M.~Whiley et al.]{I.~M.~Whiley$^{1}$\thanks{E-mail: ppximw@nottingham.ac.uk}, 
  A.~Arag{\'o}n-Salamanca$^{1}$\thanks{E-mail: alfonso.aragon@nottingham.ac.uk}, G.~De~Lucia$^{2}$, A.~von~der~Linden$^{2}$,
  \newauthor S.~P.~Bamford$^{3}$, P.~Best$^{4}$, M.~N.~Bremer$^{5}$, P.~Jablonka$^{6}$, O.~Johnson$^{4}$, 
  \newauthor B.~Milvang-Jensen$^{7,8}$, S.~Noll$^{9}$, B.~M.~Poggianti$^{10}$, G.~Rudnick$^{2}$, R.~Saglia$^{9}$, 
  \newauthor S.~White$^{2}$, D.~Zaritsky$^{11}$\\ 
$^{1}$School of Physics and Astronomy, University of Nottingham, University Park, Nottingham, NG7 2RD, UK\\ 
$^{2}$Max-Planck-Institut f{\"u}r Astrophysik, Karl-Schwarzschild-Str. 1, D-85748 Garching, Germany\\
$^{3}$Institute of Cosmology and Gravitation, Mercantile House, Hampshire Terrace, University of Portsmouth, Portsmouth, PO1 2EG, UK\\
$^{4}$SUPA, Institute for Astronomy, Royal Observatory Edinburgh, Blackford Hill, Edinburgh EH9 3HJ, UK\\
$^{5}$H H Wills Physics Laboratory, Tyndall Avenue, Bristol, BS8 1TL, UK\\
$^{6}$Universit{\'e} de Gen{\`e}ve, Laboratoire d'Astrophysique de l'Ecole Polytechnique F{\'e}d{\'e}rale de Lausanne (EPFL), Observatoire, CH--1290Sauverny, Switzerland\\
$^{7}$Dark Cosmology Centre, Niels Bohr Institute, University of Copenhagen, Juliane Maries Vej 30, DK-2100 Copenhagen, Denmark\\
$^{8}$The Royal Library/Copenhagen University Library, Research Dept., Box 2149, DK--1016 Copenhagen K, Denmark\\
$^{9}$Max-Planck-Institut f\"ur extraterrestrische Physik, Giessenbachstr., D-85741 Garching, Germany\\
$^{10}$Osservatorio Astronomico, vicolo dell' Osservatorio 5, 35122 Padova, Italy\\
$^{11}$Steward Observatory, University of Arizona, 933 North Cherry Avenue, Tucson, AZ 85721, USA 
\vspace{-0.2cm}}

\begin{document}

\date{Accepted 2008 April 10.  Received 2008 April 1; in original form 2007
November 23
\vspace{-0.3cm}
}

\pagerange{\pageref{firstpage}--\pageref{lastpage}} \pubyear{2008}

\maketitle

\label{firstpage}

\begin{abstract}  We present $K$-band data for the brightest cluster galaxies
(BCGs) from the ESO Distant Cluster Survey (EDisCS).  These data are combined
with the photometry published by \citet{Aragon-Salamanca_etal:1998} and a
low-redshift comparison sample built from the BCG catalogue of
\citet{von_der_Linden_etal:2007}. BCG luminosities are measured inside a metric
circular aperture with $37\,$kpc diameter. In agreement with previous studies,
we find that the $K$-band Hubble diagram for BCGs exhibits very low scatter 
($\sim0.35\,$mag) over a redshift range of $0<z<1$.   The colour and rest-frame
$K$-band luminosity evolution of the BCGs  are in good agreement with
population synthesis models of stellar populations which formed at $z>2$ and
evolved passively thereafter.  In contrast with some previous studies, we do
not detect any significant change in the stellar mass of the BCG since
$z\sim1$.  These results do not seem to depend on the velocity dispersion of
the parent cluster.  We also find that there is a correlation between the 1D
velocity dispersion of the clusters ($\sigma_{\rm cl}$) and the $K$-band
luminosity of the BCGs (after correcting for passive evolution).  The clusters
with large velocity dispersions, and therefore masses, tend to have brighter
BCGs, i.e., BCGs with larger stellar masses.  This dependency, although
significant, is relatively weak: the stellar mass of the BCGs changes only by
$\sim70\%$ over a two-order-of-magnitude range in cluster mass.  Furthermore,
this dependency doesn't change significantly with redshift.  We have compared
our observational results with the hierarchical galaxy formation and evolution
model predictions of \citet{de_lucia_blaizot:2007}. We find that the models
predict colours which are in reasonable agreement with the observations because
the growth in stellar mass is dominated by the accretion of old stars. However,
the stellar mass in the model BCGs grows by a factor of 3--4 since $z=1$, a growth
rate which seems to be ruled out by the observations. The models predict a
dependency between the BCG's stellar mass and the velocity dispersion (mass) of
the parent cluster in the same sense as the data, but the dependency is
significantly stronger than observed. However,  one major difficulty in this
comparison is that we have measured magnitudes inside a fixed metric aperture
while the models compute total luminosities.

\end{abstract}


\section{Introduction}

Brightest cluster galaxies (BCGs) form a very special class of galaxy. They
lie in local minima of the cluster potential well and are extremely massive
and luminous, forming their own luminosity function \citep{Sandage:1976,
Dressler:1978, Bernstein_Bhavsar:2001}.  BCGs have been the focus of many
studies (see, e.g., \citealt{Humason_etal:1956}; \citealt{Sandage:nov1972};
\citealt{Gunn_Oke:1975}; \citealt{Hoessel_etal:1980};
\citealt{Oegerle_Hoessel:1991}; \citealt{Nelson_etal:2002}) and are well known
for the low scatter in their absolute magnitudes ($\sim0.3\,$mag; see, e.g.,
\citealt{Sandage:1988}).  Early work involving BCGs took advantage of this
small scatter in absolute magnitudes, combined with the extreme luminosities
of these galaxies, to examine the cosmological bulk flow out to higher
redshifts and to try to determine the value of the deceleration parameter
$q_0$ using the BCGs as standard candles (e.g. \citealt{Sandage:nov1972};
\citealt{Sandage_Hardy:1973}; \citealt{Gunn_Oke:1975};
\citealt{Lauer_Postman:1992}).  More recent studies of BCGs have focused on
unravelling the formation and evolution of these objects in order to place
constraints on galaxy formation models. Arag\'on-Salamanca, Baugh \& Kauffmann
(1998; hereafter ABK) examined the $K$-band Hubble diagram for BCGs up to a
redshift of $\sim 1$. They found that the BCGs did not exhibit any luminosity
evolution over this redshift range, suggesting that the stellar mass of the
BCGs had grown by a factor 2--4 since $z\sim 1$, depending on the cosmology
assumed. \citet{Collins_Mann:1998} and \citet{Brough_etal:2002} also studied
the $K$-band Hubble diagram for the BCGs in an X-ray selected cluster sample.
These authors found that the amount of stellar mass growth shown by the BCGs
was dependent on the X-ray luminosity of the host cluster, with BCGs in high
X-ray luminosity clusters showing no mass accretion since $z\sim1$ and BCGs in
low X-ray luminosity clusters growing by a factor of $\sim 4$ in this time.

More recently \citet{de_lucia_blaizot:2007} have used a combination of N-body
simulations and semi-analytic techniques to study the formation and evolution
of BCGs.  They find that the formation history of local BCGs is extremely
hierarchical, with half the mass of a typical BCG being locked up in a single
galaxy after $z\sim 0.5$.  The stars that make up the BCGs are formed very
early in separate, small galaxies which then assimilate over time to form the
BCG.

In this paper we extend the work carried out by ABK, enlarging their
high-redshift sample with the BCGs from the ESO Distant Cluster Survey
(EDisCS; \citealt{White_etal:2005_alt}). The local comparison sample is also
improved using the BCGs from an SDSS-based cluster sample
\citep{von_der_Linden_etal:2007}. Moreover, additional information such as the
velocity dispersions of the host clusters is now included in the analysis.  In
section~2 we present the data and describe our main observational results
concerning the stellar mass growth of the BCGs, the evolution of their colours
and the relationship between the velocity dispersion of the host cluster and
the BCG luminosity.  Section~3 compares the observational results with the
semi-analytic model predictions of \citet{de_lucia_blaizot:2007}. Section~4
summarises our main conclusions.  We assume
$H_0=71\,$km$\,$s$^{-1}\,$Mpc$^{-1}$, $\Omega_{\rm m} = 0.27$, $\Omega_\Lambda
= 0.73$ (\citealt{Spergel_etal:2003_alt}) throughout the paper.

\section{Observational Results} 

\subsection{The sample} 

One part of the data analysed here is taken from the ESO Distant Cluster Survey
(EDisCS), presented in detail in \citet{White_etal:2005_alt}. A brief description
is given here for completeness.  The candidate clusters for EDisCS were chosen
from among the highest surface brightness objects in the Las Campanas Distant
Cluster Survey (LCDCS) of \citet{Gonzalez_etal:2001}. Potential cluster
candidates were identified in two redshift bins, mid-$z$ clusters ($0.45<z_{\rm
est}<0.55$) and high-$z$ clusters ($0.75<z_{\rm est}<0.85$), where $z_{\rm
est}$ is the estimated cluster redshift based on the magnitude of the putative
BCG (see \citealt{Gonzalez_etal:2001} for details).  The potential clusters
were checked carefully to ensure the detection appeared reliable and deep
optical and near-infrared photometry was taken for 20 fields with confirmed
cluster candidates. The optical photometry was taken using FORS2 at the VLT in
direct imaging mode.  The mid-$z$ clusters were imaged in the $B$, $V$ and $I$
bands, the high-$z$ clusters in the $V$, $R$ and $I$ bands.  Typical
integration times for the optical data were 45 minutes for the mid-$z$ clusters
and 120 minutes for the high-$z$ clusters.  The near-infrared $J$ and $K$
photometry was obtained using SOFI at the NTT (Arag\'on-Salamamca et al. 2008,
in preparation). Two fields were not imaged due to bad weather and a further
field was rejected as its spectroscopic  redshift histogram suggested the
cluster was more likely to be a chance superposition of objects spread out over
a range of redshift. However, confirmation spectroscopy  with FORS2 at the VLT
showed that several of the confirmed fields were found to contain multiple
clusters at different redshifts, which compensated for the loss of the other 3
fields \citep{Gonzalez_etal:2002,White_etal:2005_alt}. The EDisCS sample considered
here contains $K$-band data for a total of 21 BCGs spread over a redshift
range  of approximately 0.4 to 0.96.  (see Table \ref{tbl1}).  A minimum of 150
minutes of integration in the $K_{\rm s}$ band was obtained for the mid-$z$
clusters and a minimum of 300 minutes for the high-$z$ ones.  HST/ACS data was
also obtained for the 10 highest redshift clusters.  Typical random errors for
the ground based photometry are $\sim 0.007\,$mag. Errors in the photometric
zero-points are $\simeq0.02\,$mag.  The EDisCS BCGs were identified by
examining each cluster individually and choosing the most suitable galaxy in
each case, taking into account redshift, brightness and location of surface
density peak.  Further information regarding the EDisCS photometry and BCG
selection is presented in \citet{White_etal:2005_alt}.

\begin{figure}
\centering
\includegraphics[width=0.45\textwidth]{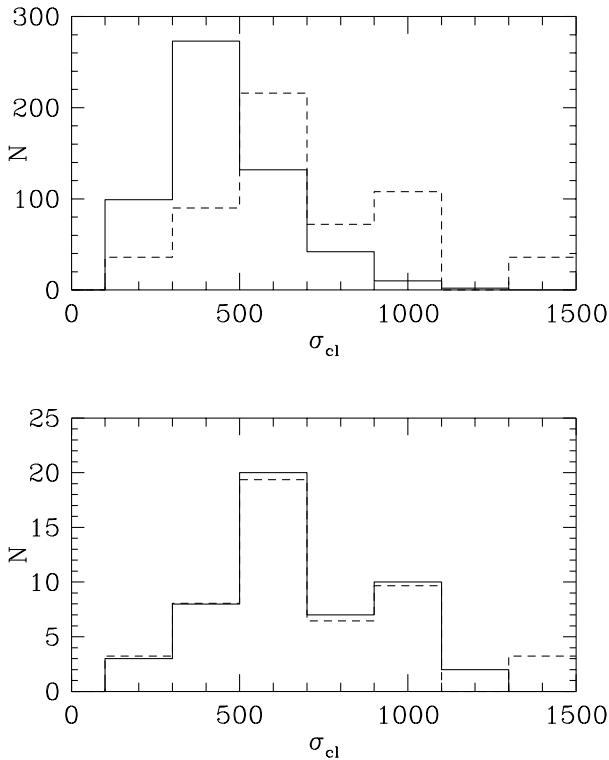}
\caption{\label{sigma_hist_plot} Top: histogram of the cluster velocity
  dispersions for the complete SDSS/2MASS (solid line) and the high-$z$
  cluster samples (dashed lines).  Bottom: as above, but using one of the
  SDSS/2MASS $\sigma_{\rm cl}$-matched subsamples.  The high-$z$ cluster
  histograms have been normalised to have the same number of objects as in the
  SDSS/2MASS samples.  }
\end{figure} 

In addition to the EDisCS BCGs we make use of a sample of local BCGs studied in
\citet{von_der_Linden_etal:2007}.  Only a brief description of how these BCGs
were selected is included here.  For a more detailed account, see
\citet{von_der_Linden_etal:2007}. The starting point of this sample was the C4
catalogue \citep{Miller_etal:2005_alt} of clusters in the Sloan Digital Sky Survey
(SDSS; \citealt{slb02_alt}, \citealt{yaa00_alt}). However, due to fiber collisions in
the SDSS (no two fibers can be placed within $55\arcsec$ of each other), about
a third of the BCGs are missed by the C4 algorithm, which is based on the
spectroscopic data alone. A semi-automatic algorithm to identify the brightest,
central galaxy of these clusters is presented in detail in
\citet{von_der_Linden_etal:2007}. Furthermore, an iterative algorithm is used
to determine the redshift $z$, velocity dispersion $\sigma_{\rm cl}$, and
virial radius $R_{200}$ from cluster members within $1 R_{200}$ from the BCG
and within $\pm 3 \sigma_{\rm cl}$ of the cluster redshift. The final cluster
sample contains 625 groups and clusters. Of their BCGs, 604 are found in the
2MASS Extended Source Catalogue (2MASSX; \citealt{Skrutskie_etal:2006_alt}).
Further cuts to the local sample during the process of obtaining reliable
$K$-band aperture magnitudes (see Section~\ref{hubble_sect}) reduced the total
number of SDSS/2MASS BCGs to 598, covering a redshift range $0.02\le z\le
0.10$.  These galaxies will be used in the following sections to build a
low-redshift comparison sample matching, as closely as possible, our
high-redshift one.

We also combine the EDisCS and SDSS/2MASS data with the photometry published
by ABK.  We refer the reader to the original paper for a description of their
data. ABK provides $K$-band photometry for a sample of 25 BCGs covering a
redshift range of $0.023<z<0.92$. These galaxies will also form part of our
study.

\begin{table*}
\begin{minipage}{126mm}
\centering
\caption{Corrected rest-frame $K$-band magnitudes for the EDisCS BCGs \label{tbl1} } 
\begin{tabular}[htb]{lcccrr}
\hline
{Cluster}      & 
{BCG ID}      & 
{$z_{cl}$}          & 
{$K_0$}        &
{$\sigma_{\rm cl}$(km/s)}$^{\rm a}$  &\\
\hline   
cl1301-1139a &  EDCSNJ1301351$-$1138356 &   0.3969  &   16.0  &   391\asymerr{69}{63}\\
cl1202-1224  &  EDCSNJ1202433$-$1224301 &   0.4240  &   16.7  &   518\asymerr{104}{92}\\
cl1037-1243a &  EDCSNJ1037523$-$1244490 &   0.4252  &   16.9  &   537\asymerr{48}{46}\\
cl1059-1253  &  EDCSNJ1059071$-$1253153 &   0.4564  &   16.0  &   510\asymerr{56}{52}\\ 
cl1018-1211  &  EDCSNJ1018467$-$1211527 &   0.4734  &   16.3  &   486\asymerr{63}{59}\\
cl1138-1133  &  EDCSNJ1138102$-$1133379 &   0.4796  &   17.0  &   732\asymerr{76}{72}\\ 
cl1301-1139  &  EDCSNJ1301402$-$1139229 &   0.4828  &   16.4  &   687\asymerr{86}{81}\\
cl1420-1236  &  EDCSNJ1420201$-$1236297 &   0.4962  &   16.1  &   218\asymerr{50}{43}\\  
cl1411-1148  &  EDCSNJ1411047$-$1148287 &   0.5195  &   16.2  &   710\asymerr{133}{125}\\ 
cl1232-1250  &  EDCSNJ1232303$-$1250364 &   0.5414  &   16.3  &   1080\asymerr{89}{119}\\
cl1037-1243  &  EDCSNJ1037514$-$1243266 &   0.5783  &   16.4  &   319\asymerr{52}{53}\\
cl1353-1137  &  EDCSNJ1353017$-$1137285 &   0.5882  &   16.5  &   666\asymerr{139}{136}\\
cl1103-1245a &  EDCSNJ1103349$-$1246462 &   0.6261  &   17.4  &   336\asymerr{40}{36}\\
cl1227-1138  &  EDCSNJ1227589$-$1135135 &   0.6357  &   17.4  &   574\asymerr{75}{72}\\
cl1054-1146  &  EDCSNJ1054244$-$1146194 &   0.6972  &   17.2  &   589\asymerr{70}{78}\\
cl1103-1245b &  EDCSNJ1103365$-$1244223 &   0.7031  &   17.1  &   252\asymerr{85}{56}\\ 
cl1040-1155  &  EDCSNJ1040403$-$1156042 &   0.7043  &   17.5  &   418\asymerr{46}{55}\\
cl1054-1245  &  EDCSNJ1054435$-$1245519 &   0.7498  &   17.5  &   504\asymerr{65}{113}\\
cl1354-1230  &  EDCSNJ1354098$-$1231015 &   0.7620  &   17.4  &   648\asymerr{110}{105}\\
cl1216-1201  &  EDCSNJ1216453$-$1201176 &   0.7943  &   17.0  &   1018\asymerr{77}{73}\\
cl1103-1245  &  EDCSNJ1103434$-$1245341 &   0.9586  &   18.0  &   534\asymerr{120}{101}\\ 
\hline
\end{tabular}

$^{\rm a}${\citet{Halliday_etal:2004_alt}, 
\citet{Milvang-Jensen_etal:2008_alt}}\\

\end{minipage}
\end{table*}

\subsection{The K-band Hubble diagram}

In order to allow a direct comparison between the data sets, we closely
followed the method adopted by ABK when constructing the $K$-band Hubble
diagram, changing only the assumed cosmological parameters.  The magnitudes of
the BCGs were measured including {\it all\/} the light contained inside a
fixed metric circular aperture centred on the BCG itself (cf.\
\citealt*{Schneider_etal:1983a}).  ABK used a metric aperture of $50\,$kpc
diameter, assuming $H_0 = 50\,$km$\,$s$^{-1}\,$Mpc$^{-1}$, and carried out
their analysis for both $q_0 = 0.0$ and $q_0 = 0.5$.  We have assumed the
cosmological parameters from \citet{Spergel_etal:2003_alt} as derived from the
Wilkinson Microwave Anisotropy Probe (WMAP), $H_0 =
71\,$km$\,$s$^{-1}\,$Mpc$^{-1}$, $\Omega_{\rm m} = 0.27$, $\Omega_\Lambda =
0.73$.  An aperture of $50\,$kpc in the $q_0 = 0.0$ cosmology assumed 
by ABK is equivalent
to an aperture of diameter $\simeq 37\,$kpc in the WMAP cosmology for the
redshifts considered here. 

For the EDisCS BCGs, the aperture magnitudes measured directly from the images
were corrected for galactic extinction using the dust maps of
\citet*{Schlegel_etal:1998}. In the $K$-band these corrections were always
small (typically $\simeq 0.01\,$mag) since the galactic latitude of the fields
is relatively high.  A $k$-correction was also applied following the same
method used by ABK using the $k$-correction published by
\citet{Mannucci_etal:2001}. The combined uncertainties in the derived $K$-band
magnitudes are $\simeq 0.05$--$0.08$, dominated by the uncertainties in the
$k$-correction (cf.\ ABK).

The method employed to measure the required aperture magnitudes for the
SDSS/2MASS BCGs is as follows.   We searched for the  positions of the 625
objects selected from the SDSS BCG sample in the  2MASS Extended Source
Catalogue (2MASSX; \citealt{Skrutskie_etal:2006_alt}).  Twenty-one of the objects
were not found, reducing the SDSS BCG comparison sample to 604 objects. 2MASSX
provided    magnitudes taken in circular apertures of increasing radius
($5\arcsec$, $7\arcsec$, $10\arcsec$, $15\arcsec$, $20\arcsec$, $25\arcsec$,
$30\arcsec$, $40\arcsec$, $50\arcsec$, $60\arcsec$ and $70\arcsec$).  The
maximum aperture for which a magnitude was available varied from object to
object.  These aperture magnitudes were then interpolated to estimate the value
corresponding to a physical aperture of $37\,$kpc at the redshift of each BCG. 
Unfortunately, for 6 of the SDSS BCGs the required aperture was larger than the
maximum aperture that was available from the 2MASSX database.  For these
objects it would have been necessary to extrapolate to estimate the required
aperture magnitudes. Rather than using uncertain extrapolated magnitudes, we
decided to discard these objects, leaving a final sample of 598 SDSS/2MASS
BCGs. Simple statistical tests show that these galaxies represent a random
subsample of the original SDSS/2MASS BCG sample, unbiased with respect to their
redshift distribution, velocity dispersion or luminosity.  Galactic extinction
corrections were applied to these objects in the same way as for the EDisCS
BCGs.  The correction was in general small (typically $\simeq 0.01\,$mag).  A
$k$-correction was also applied to the data as before.

Note that both the 2MASS and the EDisCS photometric data are in the $K_{\rm
s}$ system, while the ABK data are in the UKIRT $K$ system.  However, the
differences between the photometric systems are $\le 0.01\,$mag and thus
negligible\footnote {See the Supplement to the 2MASS All Sky Data Release,
http://www.ipac.caltech.edu/2mass/releases/allsky}.

In section~\ref{sigma_sect} we will see that the luminosity of the BCG depends
significantly on the velocity dispersion ($\sigma_{\rm cl}$) of the parent
cluster. Thus, in order to make meaningful comparisons between the low-redshift
and high-redshift BCGs, the velocity dispersions of the cluster samples need to
be reasonably well matched. The top panel of figure~\ref{sigma_hist_plot} shows
that the SDSS/2MASS sample contains a much larger proportion of low
$\sigma_{\rm cl}$ clusters than the high-$z$ sample, thus a direct comparison
cannot be made. However, given the number of clusters in the SDSS/2MASS sample,
it is possible to build reasonably-large subsamples of low-$z$ clusters with
$\sigma_{\rm cl}$ distributions that are well matched to that of the high-z
sample.  A set of 1000 local cluster subsamples, each with a $\sigma_{\rm cl}$
distribution matched to our high-$z$ sample, was built by randomly-selecting
SDSS/2MASS clusters in the right proportion. Each of the $\sigma_{\rm
cl}$-matched subsamples contains 50 SDSS BCGs. A typical realisation is shown
in the bottom panel of figure~\ref{sigma_hist_plot}. In this sample-matching
process we did not take into account the redshift evolution of $\sigma_{\rm
cl}$ because such approach would have been model dependent. However, the
expected evolution is small enough (e.g., \citealt{Poggianti_etal:2006_alt})  and
the dependence of the BCG's luminosity on $\sigma_{\rm cl}$ weak enough (cf.\
section~\ref{sigma_sect}) that our conclusions would not change if we had taken
such evolution into account.

In the analysis described below, we carried out comparisons between the
high-$z$ sample and the 1000 realisations of the low-$z$ $\sigma_{\rm
cl}$-matched sample. The quantitative results we quote always refer to the
mean of all these comparisons.  For display purposes, and for the sake of
clarity, we sometimes only plot the points corresponding to one of these
realisations whose properties are close to the mean.

\begin{figure*}
\centering
\includegraphics[width=0.70\textwidth,height=0.6\textheight]{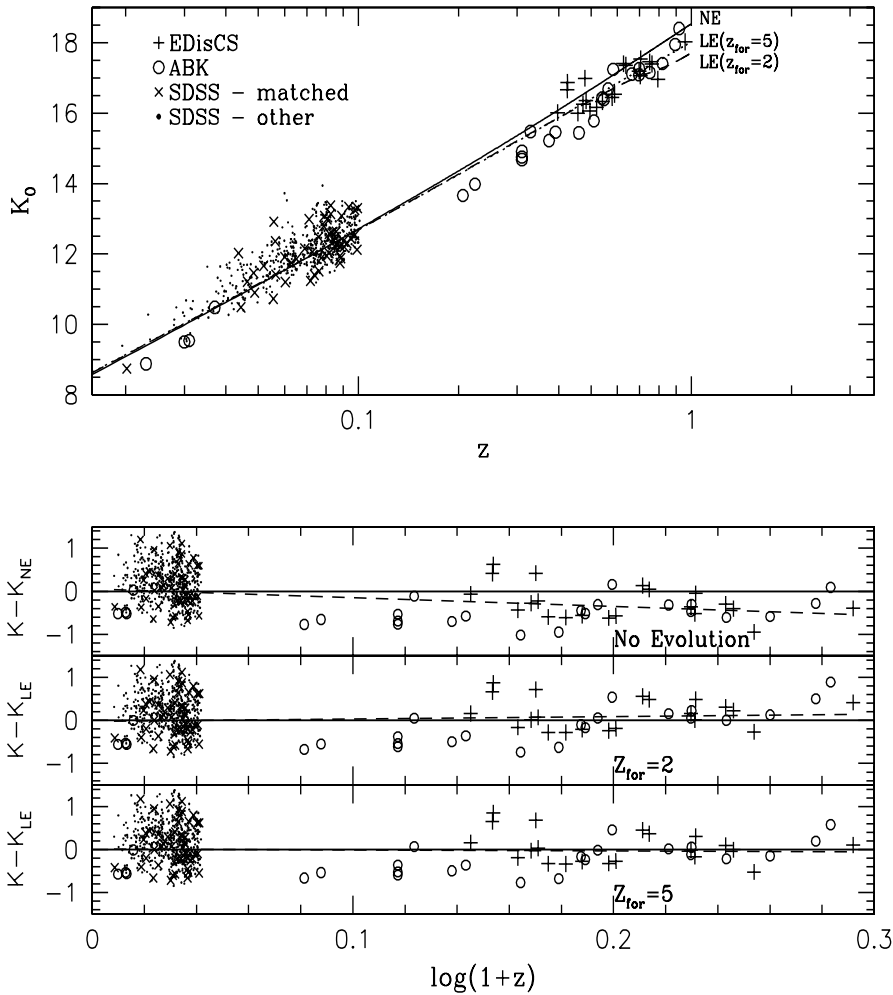}
\caption{\label{hubble_plot} Top plot: the apparent rest-frame $K$-band
  magnitudes versus redshift for the BCGs. The EDisCS points are 
  shown as plus signs, the data from ABK are represented by circles,  
  and the galaxies in one of the  
  SDSS/2MASS $\sigma_{\rm cl}$-matched subsamples
  is shown as crosses. This SDSS/2MASS subsample has properties close to the
  mean of the 1000 realisations.  
  Small dots correspond to the complete SDSS/2MASS
  BCG sample, shown here for comparison (see text for details). 
  The magnitudes have been
  measured in a fixed metric aperture of $37\,$kpc. 
  The solid line shows the no-evolution prediction, and the dashed and dotted
  lines show the predicted evolution for passively-evolving 
  stellar populations formed at
  redshifts of 2 and 5 respectively.  The model lines 
  are normalised at low-$z$ to the mean of the 1000 $\sigma_{\rm cl}$-matched 
  SDSS/2MASS subsamples.  Bottom plot: the top panel shows the data
  with the no-evolution prediction subtracted
  (symbols as above). The middle
  and bottom panels show, respectively, 
  the data after subtracting the passive evolution
  predictions for $z_{\rm form} = 2$ and $z_{\rm
  form} = 5$.  The
  dashed line in each panel indicates the average least-squares fits to each
  of the 1000 low-$z$ subsamples plus the EDisCS and ABK BCGs.}
\end{figure*}

Figure \ref{hubble_plot} shows the rest-frame $K$-band Hubble diagram  for the
BCGs in the SDSS/2MASS, EDisCS and ABK clusters.  The estimated random
uncertainties in the $K$-band magnitudes add little to the scatter of the
points, as they are $\sim0.05$--$0.08\,$mag for the EDisCS and ABK BCGs and
$\sim0.09\,$mag for the SDSS/2MASS BCGs.

\begin{table}
\centering
\caption{Corrected rest-frame $K$-band magnitudes for a typical 
  $\sigma_{\rm cl}$-matched SDSS subsample.  The properties of this
  subsample are close to the mean of the 1000 subsamples
  (See section \ref{sigma_sect}.)\label{tbl2} }
\begin{tabular}[htb]{ccrrr}
\hline
{SDSS BCG ID}      &
{$z_{\rm cl}$}          &
{$K_0\,$}        &
{$\sigma_{\rm cl}$(km/s)}$^{\rm b}$  &\\
\hline                    

587726014532550731 &    0.02015   &       8.7   &       502\asymerr{49}{49}  \\  
587722953306931328 &    0.04362   &      12.0   &       264\asymerr{34}{31}  \\  
588015507664928793 &    0.04437   &      10.5   &       933\asymerr{73}{69}  \\  
587728932421042326 &    0.04631   &      11.2   &       556\asymerr{64}{59}  \\  
587728670418272398 &    0.04845   &      11.5   &       570\asymerr{89}{77}  \\  
588848899917611022 &    0.04876   &      10.9   &       211\asymerr{56}{49}  \\  
587729387149852828 &    0.05189   &      11.7   &       547\asymerr{100}{86}  \\ 
587727227305263139 &    0.05558   &      10.7   &       903\asymerr{56}{55}  \\  
587722952231944566 &    0.05566   &      12.9   &       685\asymerr{86}{82}  \\  
587734304344178742 &    0.05625   &      12.4   &       915\asymerr{70}{67}  \\  
587733081347195107 &    0.05643   &      11.4   &       441\asymerr{45}{42}  \\  
587722983377535296 &    0.06003   &      11.9   &       506\asymerr{51}{48}  \\  
587729751131029774 &    0.06029   &      11.2   &       512\asymerr{97}{87}  \\  
587732134304350396 &    0.06288   &      11.7   &       613\asymerr{82}{77}  \\  
587731185668849907 &    0.06517   &      11.9   &      1084\asymerr{115}{113}  \\
588010360151146773 &    0.06901   &      12.2   &       800\asymerr{56}{57}  \\  
587733081882755176 &    0.07107   &      13.0   &       543\asymerr{97}{94}  \\  
587726032259973170 &    0.07174   &      11.2   &       514\asymerr{92}{84}  \\  
587732582591168714 &    0.07267   &      12.1   &       582\asymerr{108}{100}  \\
587735666921898155 &    0.07371   &      11.4   &       670\asymerr{77}{70}  \\  
587731680111820927 &    0.07587   &      12.5   &       902\asymerr{107}{96}  \\ 
587727225689538702 &    0.07618   &      11.5   &       812\asymerr{51}{48}  \\  
587726016681476254 &    0.07738   &      11.9   &       966\asymerr{60}{58}  \\  
587727212807192844 &    0.07868   &      12.8   &       765\asymerr{102}{97}  \\ 
587726031184920689 &    0.07919   &      13.1   &       327\asymerr{74}{59}  \\  
587725576423342431 &    0.08012   &      12.0   &      1156\asymerr{62}{60}  \\  
587726031184330907 &    0.08044   &      13.0   &       677\asymerr{52}{51}  \\  
588015508205994134 &    0.08157   &      12.0   &       497\asymerr{63}{62}  \\  
587725505559986486 &    0.08182   &      12.6   &       613\asymerr{92}{87}  \\  
587730817361641713 &    0.08255   &      13.1   &       406\asymerr{92}{80}  \\  
587726877271130340 &    0.08257   &      13.4   &       315\asymerr{65}{57}  \\  
587726032776593461 &    0.08316   &      12.6   &       424\asymerr{89}{78}  \\  
587733080814780480 &    0.08330   &      12.3   &       625\asymerr{74}{69}  \\  
587730816824508498 &    0.08512   &      12.5   &       229\asymerr{104}{72}  \\ 
587729782810345782 &    0.08578   &      12.3   &       530\asymerr{119}{102}  \\
588013381670994051 &    0.08581   &      12.6   &       409\asymerr{81}{73}  \\  
587729752747278348 &    0.08803   &      11.7   &       951\asymerr{57}{54}  \\  
587733410986590314 &    0.08804   &      11.9   &       705\asymerr{89}{86}  \\  
588848899917873272 &    0.08820   &      12.2   &      1020\asymerr{91}{87}  \\  
588848900454678767 &    0.08856   &      11.9   &       708\asymerr{108}{102}  \\
587733604808458243 &    0.08901   &      13.1   &       556\asymerr{110}{91}  \\ 
587736618787733576 &    0.09010   &      12.4   &       974\asymerr{61}{58}  \\  
587722982834438274 &    0.09059   &      12.2   &       748\asymerr{66}{61}  \\  
587726100415774941 &    0.09243   &      12.3   &       513\asymerr{99}{83}  \\  
587735694836039706 &    0.09356   &      13.3   &       554\asymerr{179}{162}  \\
587727212274057247 &    0.09465   &      12.4   &       513\asymerr{105}{89}  \\ 
588010359621156952 &    0.09742   &      13.2   &      1074\asymerr{105}{103}  \\
587726032234741909 &    0.09767   &      12.5   &       388\asymerr{119}{118}  \\
587729407545639093 &    0.09903   &      12.1   &      1127\asymerr{72}{71}  \\  
587726032245620919 &    0.09947   &      13.3   &       746\asymerr{203}{170}  \\
\\   
\hline
\end{tabular}

$^{\rm b}${\citet{von_der_Linden_etal:2007}}\\

\end{table}

The solid line in figure \ref{hubble_plot} is the no-evolution prediction,
which takes into account dimming solely due to distance modulus as a function
of redshift.  The other two lines plotted are passive evolution  predictions
from the population synthesis models of \citet{Bruzual_and_Charlot:2003}, using
a  \citet{Salpeter:1955}  initial mass function with solar metallicity.   We
have assumed a fixed metallicity for the BCGs as the actual metal content and
its evolution is unknown.  However, ABK showed that changing the metal content
makes little difference to the passive luminosity evolution\footnote{The
metallicity of the BCGs is discussed in section \ref{colours}.}.  The dashed
line in figure~\ref{hubble_plot} shows the predicted evolution  of a stellar
population which forms at a redshift of 2 and evolves passively thereafter. 
Similarly, the dotted line is the prediction for a passively evolving stellar
system formed at a redshift of 5.  The prediction lines are normalised at
low-$z$  to the mean of the 1000  $\sigma_{\rm cl}$-matched SDSS/2MASS
subsamples.   Contrary to the findings of ABK, who found no evidence for
luminosity evolution in the $K$-band luminosity of BCGs, figure
\ref{hubble_plot} indicates relatively small, but significant, luminosity
evolution: BCGs at $z\sim1$ appear to be, on average, 1.7~times 
brighter in the rest-frame 
$K$ band than $z\sim0$ ones. Indeed, the BCGs follow quite closely the passive
luminosity evolution lines, with a scatter of $\sim0.35\,$mag. The different
behaviour we have found is mainly due to the low-$z$ comparison sample used to
derive the low-$z$ normalisation of the Hubble diagram.  Due to the limited
amount of $K$-band data for BCGs available at the time, ABK used a small 
ad-hoc sample of only 4 low-z BCGs. The large uncertainty on the normalisation
from such a small sample was exacerbated by the fact that most of these
clusters had relatively high velocity dispersions, biasing their magnitudes.
This highlights the dangers of comparing low- and high-$z$
cluster samples without carefully matching their properties. Indeed, had we
used the entire SDSS/2MASS BCGs sample (small dots in Figure~\ref{hubble_plot})
to normalise the Hubble diagram, we would have derived a much stronger
evolution since the majority of the local clusters have low velocity
dispersions and their BCGs are therefore too faint (cf.\
section~\ref{sigma_sect}).

The three lower panels of figure~\ref{hubble_plot} show the same data after
subtracting the no-evolution and passive-evolution predictions.  As mentioned
above, it is clear from the top panel that the BCGs show a relatively small,
but significant, luminosity evolution over the redshift range plotted.   The
results of subtracting the passive evolution model predictions from the  data
are shown in the middle and bottom panels assuming $z_{\rm form} = 2$ and
$z_{\rm form} = 5$ respectively.  The lines of best fit in these panels
(derived from the average of the 1000 low-$z$ subsamples) are very close to the
passive luminosity evolution predictions themselves.  This is in agreement with
the observed colour evolution of these objects, which suggests that their stars
were formed at a redshift $z>2$ and then evolved passively (see section
\ref{colours}).

If the stellar populations of these objects are passively evolving, as their
colour evolution suggests (section \ref{colours}), the observed luminosity
evolution can be used to place limits to the amount of growth in the stellar
mass of the BCGs.  Following the method of ABK we parameterize the stellar
mass evolution as $M_{\star}(z)=M_{\star}(0)\times(1+z)^\gamma$.
Least-squares fits to the data points in the bottom panels of
figure~\ref{hubble_plot} give the following results:
$$
\matrix{
            & z_{\rm form}=2 & \ \ \ \ \gamma=-0.3\pm0.2 \cr
            & z_{\rm form}=5 & \ \ \ \ \gamma=+0.1\pm0.2 \cr
}
$$ 
This means that we do not detect any significant change in the stellar mass of
these objects. Our data is compatible with zero growth.   Formally, the derived
$\gamma$ values and their associated errors imply that, if the stellar
populations in these galaxies formed at $z_{\rm form}=2$, the stellar mass
cannot have grown by more than a factor 1.9 between $z=1$ and $z=0$ (3-$\sigma$
upper limit). For $z_{\rm form}=5$, the upper limit reduces to a factor of~1.4
(i.e., only 40\%). This contrasts with the results obtained by ABK, who
suggested stellar mass growths for the BCGs of factors between~2 and~4 since
$z\sim1$. As mentioned above, this difference is mainly due to their low-$z$
normalisation.  We compare the observed mass growth with the results from
semi-analytic models in section \ref{models}.

\begin{figure}
\centering
\includegraphics[width=0.5\textwidth]{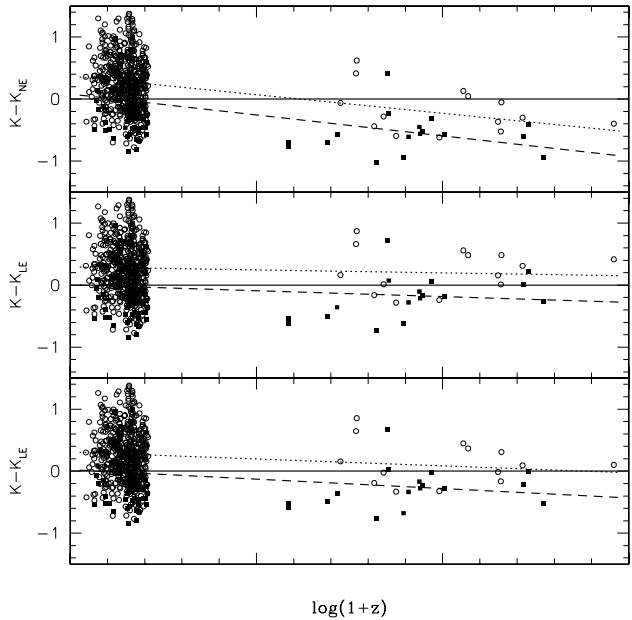}
\caption{\label{split_sigma_plot} 
The data from the $K$-band Hubble diagram  with
the model prediction lines subtracted away (cf. bottom panels of 
figure~\ref{hubble_plot}). The data has been
divided into high- and low-$\sigma_{\rm cl}$ subsamples. 
The high-$\sigma_{\rm cl}$ subsample is plotted as
solid square points and the low-$\sigma_{\rm cl}$ subsample as open 
circles. The dashed 
and dotted lines in each panel represent least-squares fits to the 
high-$\sigma_{\rm cl}$ and low-$\sigma_{\rm cl}$
data respectively. See text for details.}
\end{figure} 

\citet{Collins_Mann:1998} and \citet{Brough_etal:2002} found that the amount
of stellar mass growth shown by the BCGs could be dependent on the X-ray
luminosity of the host cluster, with BCGs in high X-ray luminosity clusters
showing little or no stellar mass growth since $z\sim1$, while BCGs in low
X-ray luminosity clusters exhibit a large variation in their stellar mass
growth, from clusters showing no growth to a few growing by factors of up to
$\sim 4$. As discussed in \S\ref{sigma_sect}, the cluster X-ray luminosity and
velocity dispersion $\sigma_{\rm cl}$ can be taken as a proxies for cluster
mass. If we divide our low- and high-$z$ cluster samples into two equal
subsamples, one with $\sigma_{\rm cl}\le654\,$km$\,$s$^{-1}$ and the other
with $\sigma_{\rm cl}>654\,$km$\,$s$^{-1}$ we find that the inferred stellar
mass growth does not seem to depend on $\sigma_{\rm cl}$: the stellar mass of
the BCGs seems to have remained roughly constant for both subsamples (albeit,
with weaker formal limits given the smaller sample sizes).
This is shown in figure~\ref{split_sigma_plot}, which presents again 
the lower panels of figure~\ref{hubble_plot}
with the data split into high and low $\sigma_{\rm cl}$ subsamples. 
There is no significant difference in the luminosity evolution rate of 
the two $\sigma_{\rm cl}$ subsamples. 
However, there is clearly an offset between the lines
of best fit in each panel of figure~\ref{split_sigma_plot}, 
with the high-$\sigma_{\rm cl}$ sample being brighter
in each case. This implies a
significant dependence of the
luminosity of the BCG on cluster velocity dispersion, as 
discussed in section~2.4.

To test whether our conclusions depend strongly on the choice of galaxy
evolution models, we have re-calculated the expected passive $K$-band
luminosity evolution (cf. figure~\ref{hubble_plot}) using  the models of
\citet{Maraston_et_al:2006}. Since these models use a very different
prescription for TP-AGB stars, the near-IR  luminosity evolution is
significantly different from that of \citet{Bruzual_and_Charlot:2003} for
stellar populations with ages $\sim0.5$--$1\,$Gyr. However, for the range of
ages relevant to our analysis ($2.6$--$12.5\,$Gyr), the model predictions agree
within $\simeq0.1\,$mag for a given metallicity and IMF. We are thus confident
that the conclusions of this section are reasonably model-independent.

\label{hubble_sect}

\subsection{Colours}

\begin{figure}
\centering
\includegraphics[width=0.5\textwidth]{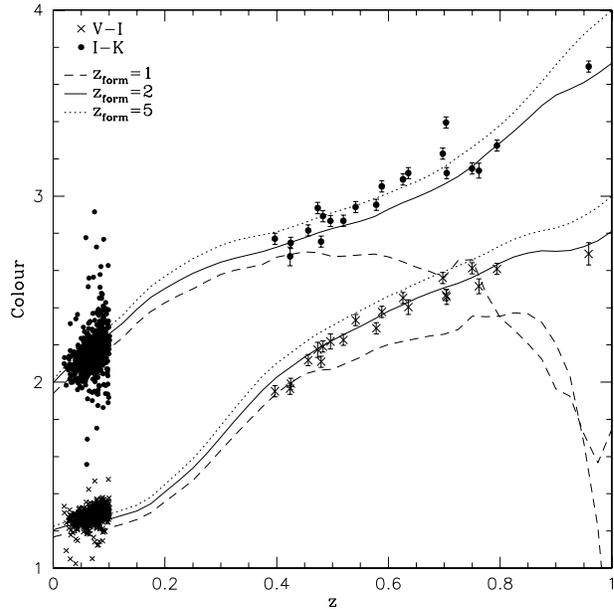}
\caption{\label{colour_plot} Plot of $V-I$ (crosses) and $I-K$ (dots)
  observer-frame colours for the EDisCS BCGs ($z > 0.4$) and the SDSS/2MASS 
  BCGs ($z < 0.4$).  
  The lines are predictions from the population synthesis code of
  \citet{Bruzual_and_Charlot:2003}.  The solid lines represent the colour
  evolution prediction for a stellar population formed at a redshift of 2.
  The dotted and dashed lines are the predictions for stellar populations with
  $z_{\rm form} = 5$ and $z_{\rm form} = 1$ respectively.  The predictions
  were computed using a \citet{Salpeter:1955} IMF and solar
  metallicity.}
\end{figure}

Figure \ref{colour_plot} shows the observed-frame $V-I$ and $I-K$ colours for
the EDisCS and SDSS/2MASS BCGs, plotted as a function of redshift.   For the
EDisCS BCGs a $2\arcsec$ radius circular aperture was used to measure the
colours. This aperture is large enough to sample a significant fraction of the
BCG's light, while keeping sky-subtraction errors small\footnote{Measuring
galaxy  colours inside the metric aperture used in section~\ref{hubble_sect} 
(average radius$\sim2.5\arcsec$) makes no significant difference to  the
conclusions of this section.}.  The $2\arcsec$ radius at the average redshift
of the EDisCS BCGs, ($\sim0.6$),  corresponds to an aperture of radius
$\sim10\arcsec$ at the average redshift of the SDSS BCGs ($\sim0.07$). Optical
and near-infrared  photometry for the SDSS/2MASS sample was measured inside
circular apertures  with $10\arcsec$ radius, and transformed into the $VIK$
system following \cite{Blanton_and_Boweis:2007}.

The colours were corrected for galactic extinction but not $k$-corrected.  The
lines shown in figure \ref{colour_plot} are predictions created using the
population synthesis code of \citet{Bruzual_and_Charlot:2003}.  The solid lines
represent the expected colour evolution of a stellar population with solar 
metallicity and a \citet{Salpeter:1955} IMF which formed in an  instantaneous
burst at $z = 2$ and then evolved passively.   The dotted and dashed lines show
the expected evolution for similar stellar populations that formed at $z = 5$
and $z = 1$ respectively.  It is clear from figure \ref{colour_plot} that,
within the model uncertainties (see below), there is reasonably good agreement
between the data and the passive evolution  models with $z_{\rm form} = 2$ and
5.  As expected, the model for $z_{\rm form} = 1$ does not fit the data at
all.  The agreement between the colours of the BCGs and the passive evolution
models is a strong indication that the BCGs are composed of stellar populations
that formed at $z \ga 2$ and have been evolving passively since. Similar
conclusions are obtained using the  \citet{Maraston_et_al:2006} models. For the
range of ages relevant  to $z_{\rm form}$ between~2 and~5, these models predict
colours which agree with those of \citet{Bruzual_and_Charlot:2003} within
$\simeq0.1\,$mag. This should be taken as an indication of the typical
uncertainties associated with  model colours. For that reason, the fact that
the $V-I$  BCG colours seem to agree with the  $z_{\rm form}=5$ models while
for $I-K$ there is better agreement  with $z_{\rm form}=2$ is not
significant. Nevertheless, since the $V$ band samples the rest-frame
ultraviolet, a very small amount of recent star formation (say, $\sim1$\% by
mass) could easily make the observed $V-I$ colours bluer by $\sim0.1$ while
having negligible effect on $I-K$.

As the metallicity of the BCGs is unknown, the colour evolution predictions
were also calculated for a stellar population with approximately double solar
metallicity and also approximately half solar metallicity.  These predictions
were found to lie nowhere near the plotted data points, being far too red and
blue respectively to reproduce the colour evolution exhibited by the BCGs.
This shows that although the colour evolution predictions are sensitive to
changes in metallicity, the assumption of approximately solar metallicity is a
reasonable one.

\label{colours}

\subsection{BCG $K$-band luminosity and cluster velocity dispersion.}

\label{sigma_sect}  

Arguably, the most fundamental property of a cluster is its mass, so it is not
unreasonable to expect that the luminosity (and stellar mass) of the BCGs may
be related to it. Several observational properties can be used as proxies for
the cluster mass, including the cluster X-ray luminosity and temperature and
the cluster velocity dispersion $\sigma_{\rm cl}$. Indeed,
\cite{Edge_and_Stewart:1991} found a significant correlation, in a low-$z$
sample, between the optical luminosity of the BCGs and the X-ray luminosity
and temperature of the parent clusters in the sense that hotter and more X-ray
bright clusters host brighter BCGs.

Velocity dispersions are available for all the EDisCS and SDSS clusters and 10
of the clusters studied by ABK.  Columns 4 of tables \ref{tbl1} and \ref{tbl2}
show the velocity dispersions $\sigma_{\rm cl}$ of the EDisCS and SDSS/2MASS
$\sigma_{\rm cl}$-matched cluster samples.  A full description of the
calculation of the $\sigma_{\rm cl}$ values for the SDSS clusters can be found
in \citet{von_der_Linden_etal:2007}.  Information regarding the EDisCS
$\sigma_{\rm cl}$ values can be found in \citet{Halliday_etal:2004_alt},
\citet{Poggianti_etal:2006_alt} and \citet{Milvang-Jensen_etal:2008_alt}.
The velocity dispersions for the ABK clusters are taken from
\citet{Girardi_Mezzetti:2001} and \citet{Lubin_etal:2002}. Note that in all
these studies the cluster velocity dispersions have been determined following
very similar methods, and are thus directly comparable.

In order to test for any relationship between the BCG $K$-band luminosity (or
stellar mass) and the cluster mass, we have calculated the residuals around the
three lines of best fit shown in the bottom three panels of
figure~\ref{hubble_plot} and plotted them against $\log_{10}(\sigma_{\rm cl})$
in figure \ref{sigma_plot}.  This procedure should correct for redshift-related
effects.  The panels show, from top to bottom, the residuals around the line of
best fit after the rest-frame $K$-band magnitudes have been corrected for no
evolution, passive evolution with $z_{\rm form} = 2$ and passive evolution with
$z_{\rm form} = 5$, respectively (cf. figure~\ref{hubble_plot}). The symbols
for the EDisCS and the ABK  samples are the same as in 
figure~\ref{hubble_plot}.
As an illustration, we also  show (crosses) the BCGs in the SDSS/2MASS
$\sigma_{\rm cl}$-matched subsample whose line of best fit (see below) is
closest to  the mean of the 1000 subsamples.

There is a clear trend in each panel of figure~\ref{sigma_plot} indicating
that clusters with large velocity dispersions, and presumably large masses,
tend to have brighter BCGs.  A Spearman rank correlation
coefficient analysis indicates that all these 
correlations are significant at a level greater than 99.9\%. 

To explore whether the luminosity-$\sigma_{\rm cl}$ correlation evolves 
with redshift, we analysed the
intermediate-redshift EDisCS/ABK clusters and the low-redshift SDSS ones
separately.  We define $L_K^{\rm c}$ as the rest-frame $K$-band luminosity
of the BCGs corrected either assuming 
no luminosity evolution (N.E.) or passive luminosity evolution for 
$z_{\rm form}=2$ or~5.  
If we parameterize the trends shown in figure~\ref{sigma_plot} as
$L_K^{\rm c}=C\sigma_{\rm cl}^\alpha$, simple least-squares fits to the
EDisCS/ABK data give
$$
\matrix{
            & N.E.           & \ \ \ \  \alpha=0.35 \pm 0.14\phantom{,}\cr
            & z_{\rm form}=2 & \ \ \ \  \alpha=0.36 \pm 0.14\phantom{,}\cr
            & z_{\rm form}=5 & \ \ \ \  \alpha=0.34 \pm 0.14,\cr
}
$$ and for the mean of the 1000 SDSS/2MASS $\sigma_{\rm cl}$-matched cluster
subsamples we get $\alpha=0.35 \pm 0.10$ in all cases. 

Note that the correlations found and their estimated slopes 
are robust against changes in the way we
account for the passive luminosity evolution of the BCGs.

The slopes of the $L_K$--$\sigma_{\rm cl}$
relations are the same  for the low- and high-$z$ samples. This seems to
indicate that in the the range of redshifts explored here the rate of the BCG
buildup and that of the cluster are linked in a way that does not depend
strongly on cosmic time.

Another interesting result is that, although significant, the dependency of
$L_K^{\rm c}$ (presumably proportional to the BCG's stellar mass $M_{\rm
BCG}^*$) on $\sigma_{\rm cl}$ is relatively weak. If we assume  that the
cluster mass $M_{\rm cl}$ is proportional to $R_{200}\sigma_{\rm cl}^2$, where
$R_{200}$ is the radius of a sphere with interior mean density 200 times the
critical density of the universe at the cluster redshift. If we also assume
that, at a given redshift, $R_{200}\propto\sigma_{\rm cl}$
(\citealt{Carlberg_et_al:1997}; \citealt{Finn_et_al:2005_alt}), then $M_{\rm
cl}\propto \sigma_{\rm cl}^3$, and $M_{\rm BCG}^*\propto M_{\rm
cl}^{0.12\pm0.03}$. Thus, the stellar mass of BCGs changes only by $\sim70\%$
over a two-order-of-magnitude range in cluster mass. This is not surprising: it
has long been known that BCG luminosities measured inside fixed metric apertures
provide reasonable standard candles almost independent of the cluster richness
(\citealt{Postman_and_Lauer:1995}). 

Very recently, \citet{Brough_etal:2008} and  \citet{Stott_et_al:2008} have
found similar results for X-ray selected cluster samples.

\section{Comparison with galaxy formation models}

\begin{figure}
\centering
\includegraphics[width=0.45\textwidth]{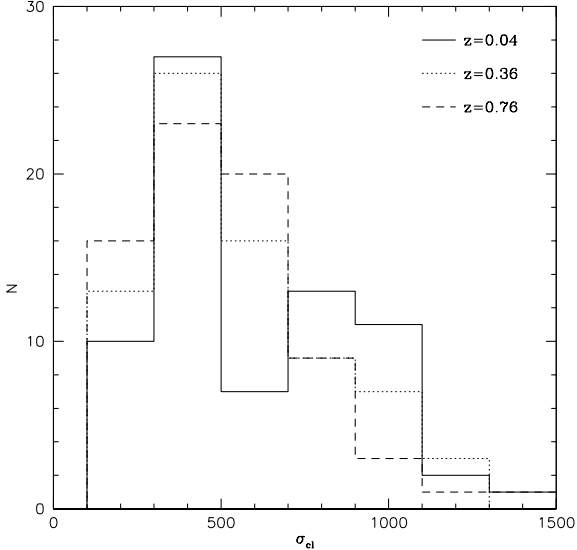}
\caption{\label{model_sigma_hist_plot} 
Histograms showing the cluster velocity dispersion $\sigma_{\rm cl}$ 
distribution
of the model clusters. The solid line shows the distribution of the 
model $\sigma_{\rm cl}$
values at $z = 0.04$. The dotted and dashed lines 
show the $\sigma_{\rm cl}$ distributions
at $z = 0.36$ and $z = 0.76$ respectively. 
The $\sigma_{\rm cl}$ values were selected using the second
method described in the text with $\sigma_{\rm cl}=200\,$km$\,$s$^{-1}$.
}
\end{figure} 

In this section, we compare our empirical results with theoretical predictions
from \citet{de_lucia_blaizot:2007}.  These authors use a combination of large
N-body collisionless cosmological simulations and semi-analytic techniques.
The semi-analytic models are grafted onto the hierarchical merging trees
extracted from the Millennium Simulation \citep{Springel_etal:2005_alt}.  In this
study, we compare our results with two versions of the semi-analytic model
which differ for the prescriptions adopted for the supernovae feedback.  In
the following we will refer as model 1 to a model which employ the same
supernovae feedback model used in \citet{de_lucia_etal:2006} where the
ejection rates are computed on the basis of energy conservation arguments.  We
will refer as model 2 to a model which adopts the same supernovae feedback
model as in \citet{Croton_etal:2006_alt} where the ejection rates are proportional
to the star formation rate.  \citet{de_lucia_blaizot:2007} have shown that
model 1 results in a slower evolution of the K-band magnitudes as a function
of redshift, albeit producing K-band magnitudes of local BCGs brighter than a
factor of about $0.8$~mag with respect to results from model 2.

In order to compare our results to model predictions, we extracted $90$ haloes
within the simulation box, uniformly distributed in log(mass) between $5\times
10^{12}\,{\rm M}_{\sun}$ and $5\times10^{15}\,{\rm M}_{\sun}$.  We extracted
the haloes from the output of the simulation at $z=0$ and followed their
history back in time by tracking, at each time, the main progenitor branch
defined as in \citet{de_lucia_blaizot:2007}.  For each SN feedback model, we
have then created tables with rest-frame and observed-frame magnitudes in 5
different bands ($B$, $V$, $R$, $I$, $J$, and $K$) for the brightest galaxy in
each halo.  For each halo we have also calculated the 3D and projected
velocity dispersions using all galaxies within $2\times{\rm R}_{200}$ at three
redshifts ($z=0.04$, $0.36$, and $0.76$).  In order to simulate the errors
associated with the observed $\sigma_{\rm cl}$ values, for each halo and each
redshift we extracted a random number of objects, between $20$ and $70$ from
all galaxies within $2\times{\rm R}_{200}$ and computed the 1D velocity
dispersion and their confidence limits using the same method that was used to
calculate the velocity dispersions for the EDisCS clusters
\citep{Halliday_etal:2004_alt}\footnote{In the smallest haloes there are cases
where the number of objects is less than 20.  In these cases we take all
objects for the calculation of the velocity dispersion.  Note, however, that
this situation seldom arises for the final comparison sample, once a minimum
velocity dispersion cut is applied (as we do in our analysis).}. This method
tries to mimic the observational uncertainties in the model predictions.

When selecting the model cluster sample to compare with observations, there
are two obvious alternatives. First, one could follow the evolution of the
same model clusters, i.e., select clusters at all redshifts which, at a given
redshift (e.g., $z=0$), have $\sigma_{\rm cl}>\sigma_{\rm lim}$. Second, one
could follow similar cluster ``samples'', i.e., select clusters that, at each
redshift, have $\sigma_{\rm cl}>\sigma_{\rm lim}$. Obviously, the first
selection tells us what happens to the same set of galaxies, while the second
one is probably closer to observations since it follows comparable samples. We
will therefore follow the second selection method with $\sigma_{\rm
lim}\simeq200\,$km$\,$s$^{-1}$ (64--70 model clusters, depending on redshift)
and $\sigma_{\rm lim}\simeq400\,$km$\,$s$^{-1}$ (28--45 model clusters, also
depending on redshift), which seem reasonable limits for comparison with our
dataset (cf. figure~\ref{sigma_hist_plot}). Note that 
the $\sigma_{\rm lim}$ for EDisCS is not 
well defined since velocity dispersion was not explicitly part of 
the EDisCS cluster selection 
criteria. However, since the results do not depend on the precise
value of this limit, the choice is not critical. 

Figure~\ref{model_sigma_hist_plot} shows the model  $\sigma_{\rm cl}$
distribution at each redshift bin obtained using this selection technique. 
Although the model $\sigma_{\rm cl}$ values were not explicitly  matched in
each redshift bin, as was done for the observational data, their distributions
are well matched at all redshifts considered. Moreover, they are also well
matched to the $\sigma_{\rm cl}$ values of the observational data 
(c.f.~figure~\ref{sigma_hist_plot})

\begin{figure}
\centering
\includegraphics[width=0.45\textwidth]{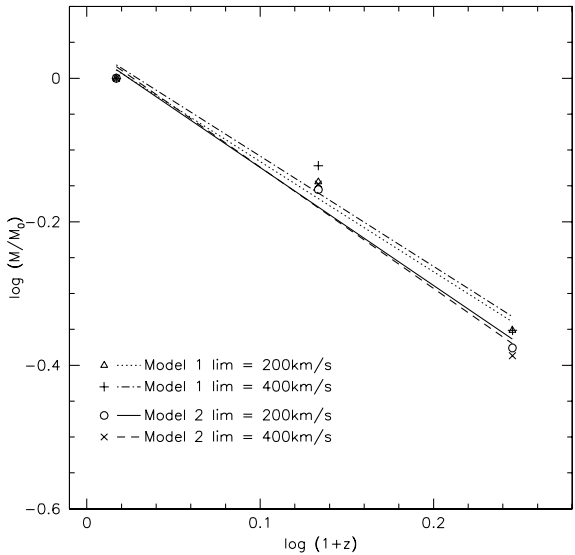}
\caption{\label{mass_growth_plot} 
Plot of $\log(M/M_0)$ versus $\log(1 + z)$ for the semi-analytic models. 
$M_0$ is the mean stellar mass of the BCGs at 
$z = 0.04$, and $M$ the mean stellar mass of the BCGs at different redshifts.
The triangles represent the
data from model 1 with $\sigma_{\rm lim} = 200\,$km$\,$s$^{-1}$. 
The doted line is a least-squares fit to
these data points. See legend for the meaning of the rest of the symbols and
lines. 
}
\end{figure}

The first result of the comparison is that the colours of the model galaxies
are similar to the observed ones for model~1, i.e., close to those of
passively-evolving stellar populations formed early.  This suggests that even
if the buildup of the galaxies' stellar mass is gradual, this growth is
dominated by the accretion of old stellar populations rather than by new star
formation.  Model~2 predicts somewhat bluer colours.

Both models 1 and 2 predict a growth in stellar mass that is significantly
stronger than observed: the stellar mass of model BCGs grows by about a factor
of $\sim3$--$4$ between $z\sim1$ and $z\sim0$, while the observations are
compatible with no growth at all and seem to rule out a factor of two growth
since $z\sim1$ (cf. section~\ref{hubble_sect}). The stellar mass growth in the
models does not depend strongly on which feedback method is used, although it
is marginally stronger in model 2. The stellar mass growth is also quite
insensitive to the value of $\sigma_{\rm lim}$, which agrees with the observed
independence of the stellar mass growth on $\sigma_{\rm cl}$. These results are
illustrated in figure~\ref{mass_growth_plot}.

\begin{figure*}
\centering
\includegraphics[width=0.7\textwidth]{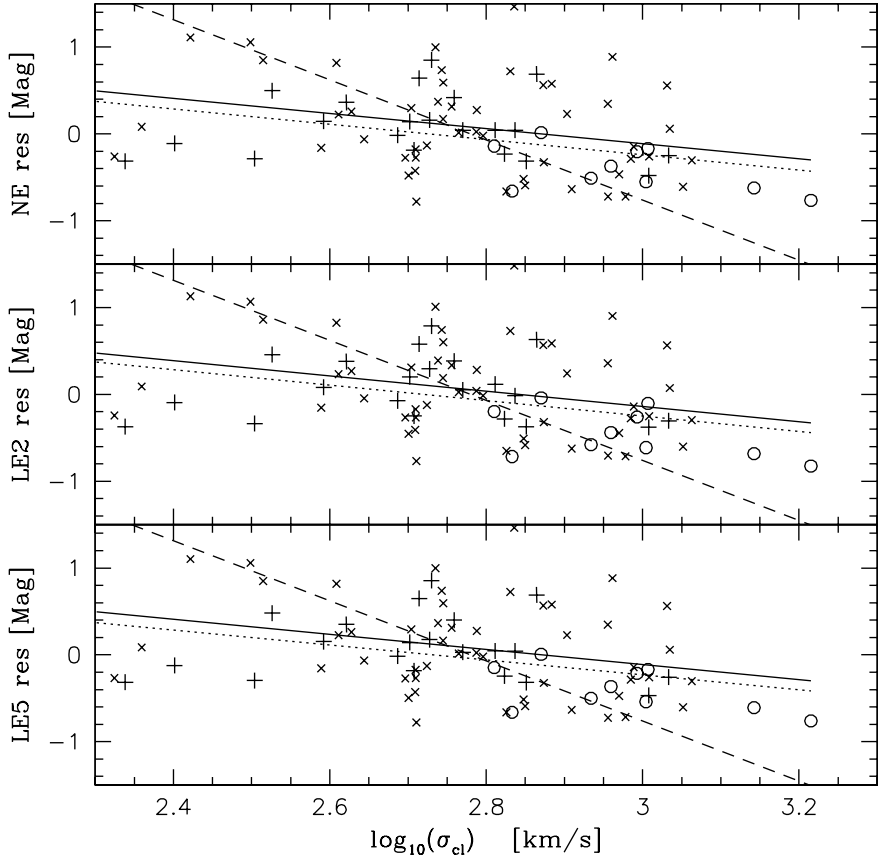}
\caption{\label{sigma_plot} Plots of the residuals from the lines of
  best fit in figure \ref{hubble_plot} vs.\ $\log_{10}(\sigma_{\rm
  cl})$.  The
  top panel shows the residuals after subtracting the no evolution
  prediction. The middle and bottom panels show the same but for the passive
  evolution models with $z_{\rm form} = 2$ and $z_{\rm form} = 5$
  respectively.  The EDisCS clusters have been plotted as plus signs, the
  points from the ABK sample for which we have velocity dispersions are
  plotted as circles and the data for a typical SDSS/2MASS $\sigma_{\rm
  cl}$-matched subsample have been plotted as crosses.  
  The SDSS/2MASS subsample shown
  has a slope very close to the mean slope of all 1000 subsamples
  (see text for details).  
  The dotted lines represent the least-squares fits to
  the EDisCS/ABK data and the solid lines represent the average least-squares
  fits to the 1000 SDSS/2MASS subsamples.  The
  dashed lines represent the average slope predicted by the
  semi-analytic models described in the text.  }
\end{figure*}

Finally, we compare the observed dependency of the BCGs' corrected $K$-band
luminosity (or stellar mass) with the cluster velocity dispersion (cf.
section~\ref{sigma_sect}). For model~1, regardless of the exact value of
$\sigma_{\rm lim}$, the models predict, at all the redshifts considered,
$L_K^{\rm c}\propto\sigma_{\rm cl}^{1.4}$ or $M_{\rm BCG}^*\propto M_{\rm
cl}^{0.5}$.  Model~2 predicts $L_K^{\rm c}\propto\sigma_{\rm cl}^{1.2}$ or
$M_{\rm BCG}^*\propto M_{\rm cl}^{0.4}$.  The data suggest a much weaker
dependency of the stellar mass on the cluster velocity dispersion or
mass. That is clearly shown in figure~\ref{sigma_plot}: the dashed line
represents the average slope from the model predictions, which is
significantly steeper than the data.

One major caveat in the comparison of our observations with the semi-analytical
model results is that we have used magnitudes measured inside a fixed metric
aperture while the models compute total magnitudes. Observationally it is very
difficult (virtually impossible) to measure accurate total magnitudes for the
BCGs  (see, e.g., \citealt{Gonzalez_etal:2005}). Indeed, it is not even clear
whether one should include the intracluster light (ICL) as part of the BCG's
luminosity (\citealt{Gonzalez_etal:2007}). A  $37\,$kpc diameter aperture
encompasses only 25\%--50\% of the total  light contained in the BCG and ICL
(Gonzalez, Zaritsky and Zabludoff, private communication). Interestingly, 
using the $I$-band data published by these authors we find
that although their $37\,$kpc aperture magnitudes show a dependency with 
$\sigma_{\rm cl}$ compatible with our measurements,  their total
BCG+ICL  luminosities show a significantly steeper dependence  ($L_I^{\rm
BCG+ICL}\propto\sigma_{\rm cl}^{0.6}$). Although the slope is still
considerably shallower than  the model predictions, this could explain at least
part of the disagreement between models and observations. Note, however, that
the scatter is large and their sample relatively small.

On the other hand, the models do
not have spatial information regarding the distribution of the BCG light, so
aperture magnitudes cannot be calculated.  It is also not clear whether the
luminosities of the model BCGs contain some of what, observationally, we could
call intracluster light. The luminosity of the model BCG includes the stars
formed in the central cluster halo and the stars formed in galaxies that
merged with the central galaxy, but not stars stripped from other cluster
galaxies due to tidal and ``harassment'' effects, since these are not modeled.
Therefore, any discrepancy (and/or agreement) found in the behaviour of the
observed and model BCGs should be taken with caution, since not all the physical
effects which potentially contribute to the growth of the BCGs are taken into
account.

A possible way of reconciling our observations and the model predictions could
be that, if there is a large stellar mass growth in BCGs (e.g.,  a factor of 4
since $z\sim1$), these stars could   lie outside of the $37\,$kpc apertures we
have used in our measurements. In that case, such a growth would really
represent the growth of the intracluster component rather than the central BCG.
How such growth would depend on the cluster properties is difficult to
predict, but it is clear that the intracluster component could play a very 
important r\^ole in the formation and evolution of clusters  
(see, e.g., \citealt{Lin_and_Mohr:2004}). 
For recent theoretical studies of the formation of the intracluster 
light, see~\citet{Monaco_et_al:2006} and~\citet{Conroy_et_al:2007}.

\label{models}

\section{Conclusions}

In this paper we present new $K$-band data for 21 brightest cluster galaxies
(BCGs) from the ESO Distant Cluster Survey (EDisCS).  Combining these  data
with the photometry published by \citet*{Aragon-Salamanca_etal:1998} we have 
built a sample of 42 BCGs with good quality near-infrared photometry in the 
$0.2<z<1$ redshift range. We have also put together a low-redshift ($\langle
z\rangle=0.07$) comparison sample from the BCG catalogue of
\citet{von_der_Linden_etal:2007} for which optical and near-infrared photometry
is available from the SDSS and 2MASS. By comparing the properties of the BCGs
in the low- and high-$z$ datasets and those of their parent clusters we have
carried out a detailed study of the evolution of these galaxies. Since, as we
demonstrate, the properties of the BCGs depend upon the velocity dispersion 
$\sigma_{\rm cl}$ (and thus presumably the mass) of the cluster they inhabit,
we carefully match the $\sigma_{\rm cl}$ distributions of our samples before
comparing them. In this work the luminosities of the BCGs are measured inside
a fixed metric circular aperture with $37\,$kpc diameter.  
The main conclusions of our study are:

\begin{enumerate}

\item In agreement with previous studies,  we find that the $K$-band
Hubble diagram for BCGs exhibits very low scatter  ($\sim0.35\,$mag) over a
redshift range of $0<z<1$.   

\item The colour evolution of BCGs as a function of redshift  shows good
agreement with population synthesis models of stellar populations that formed
at $z>2$ and evolved passively thereafter.

\item In contrast with some previous studies  (e.g., Arag\'on-Salamanca et al.\
1998) we  find that the rest-frame $K$-band luminosity evolution of the BCGs is
also well reproduced  by passive evolution models. The main reason for this
discrepancy is the use of a different, hopefully more accurate, low redshift
comparison sample with well-matched $\sigma_{\rm cl}$ distribution.

\item The agreement of the observations with passive luminosity evolution
models means that we do not detect any significant change in the stellar mass
of the BCG since $z\sim1$. Formally, we rule out a growth in the stellar mass
larger than a factor of 1.9 between $z=1$ and $z=0$ (3-$\sigma$ upper limit). 
This result does not seem to depend on the velocity dispersion of the parent
cluster.

\item We find that there is a correlation between $\sigma_{\rm cl}$  (the 1D
velocity dispersion of the clusters)  and the $K$-band luminosity of the BCGs
(after correcting for passive evolution).  Clusters with large velocity
dispersions, and therefore masses, tend to have brighter BCGs, i.e., BCGs with
larger stellar masses.  This dependency, although significant, is relatively
weak: the stellar mass of the BCGs changes only by $\sim70\%$ over a
two-order-of-magnitude range in cluster mass.  Furthermore, this dependency
doesn't change significantly with redshift.  

\end{enumerate}

We have compared our observational results with the hierarchical galaxy
formation and evolution model predictions of \citet{de_lucia_blaizot:2007}. 
One major difficulty in this comparison is that we have measured magnitudes 
inside a fixed metric aperture while the models compute total
luminosities. We find that the models predict colours which are in reasonable
agreement with the observations because the growth in stellar mass is dominated
by the accretion of old stars. However, the stellar mass in the model BCGs
grows by a factor of $\sim3$--$4$  since $z=1$, a growth rate which seems to be
ruled out by the observations. The models predict a dependency between the
BCG's stellar mass and the velocity dispersion (mass) of the parent cluster in
the same sense as the data, but the dependency is significantly stronger than
observed. It therefore seems as if building realistic BCGs is another challenge
for this kind of models, indicating that the underlying physics is far from
being completely understood.

\section*{Acknowledgments}

This paper is partly based on observations collected at the European Southern
Observatory, Chile, as part of large programme 166.A--0162 (the ESO Distant
Cluster Survey). This publication makes use of data products from the Two
Micron All Sky Survey, which is a joint project of the University of
Massachusetts and the Infrared Processing and Analysis Center/California
Institute of Technology, funded by the National Aeronautics and Space
Administration and the National Science Foundation. IW and AAS acknowledge
financial support from PPARC.  The Dark Cosmology Centre is funded by the
Danish National Research Foundation.

This publication makes use of data products from the Sloan Digital Sky Survey
(SDSS).  Funding for the SDSS and SDSS-II has been provided by the Alfred P.
Sloan Foundation, the Participating Institutions, the National Science
Foundation, the U.S. Department of Energy, the National Aeronautics and Space
Administration, the Japanese Monbukagakusho, the Max Planck Society, and the
Higher Education Funding Council for England. The SDSS Web Site is
{\tt http://www.sdss.org/}.

The SDSS is managed by the Astrophysical Research Consortium for the
Participating Institutions. The Participating Institutions are the American
Museum of Natural History, Astrophysical Institute Potsdam, University of
Basel, University of Cambridge, Case Western Reserve University, University of
Chicago, Drexel University, Fermilab, the Institute for Advanced Study, the
Japan Participation Group, Johns Hopkins University, the Joint Institute for
Nuclear Astrophysics, the Kavli Institute for Particle Astrophysics and
Cosmology, the Korean Scientist Group, the Chinese Academy of Sciences
(LAMOST), Los Alamos National Laboratory, the Max-Planck-Institute for
Astronomy (MPIA), the Max-Planck-Institute for Astrophysics (MPA), New Mexico
State University, Ohio State University, University of Pittsburgh, University
of Portsmouth, Princeton University, the United States Naval Observatory, and
the University of Washington.

\bibliographystyle{mn2e} \bibliography{papers_cited_by_AAS} \bsp

\label{lastpage}
\end{document}